      [1999/12/01 v1.4c Il Nuovo Cimento]
\documentclass{cimento}

\usepackage{graphicx}  
\usepackage{amsmath}
\title{Models for transverse-momentum distributions and transversity}
\author{A.~Bacchetta\from{ins:x}}
\instlist{\inst{ins:x} Dipartimento di Fisica Nucleare e Teorica,
  Universit\`a di Pavia, and INFN, Pavia,
\\
via Bassi 6, Pavia, Italy}

\begin{document}

\maketitle

\begin{abstract}
I present a short
review of models for transverse-momentum distributions and transversity, 
with a particular attention on general features common to many models. I
compare some model results with experimental extractions. 
I discuss the existence of
relations between different functions, their limits of validity, their
possible use.
\end{abstract}

\section{Introduction}

We believe that the dynamics of quarks and gluons inside the nucleon is
described by Quantum ChromoDynamics (QCD). However, due to the difficulty of
solving QCD in the nonperturbative regime, we are unable to calculate 
a priori most of the properties of the nucleon structure, nor to demonstrate
directly that QCD leads to parton confinement (see the talk by
D.~Sivers~\cite{Sivers:2011ci}).  
The only exception are the
studies performed with lattice QCD, which however still present several
limitations~\cite{Hagler:2009ni}.  

In this context, it may be useful to resort to effective models that try to
describe the structure of the nucleon using some reasonable
simplifications. Models cannot replace full calculations based on the
underlying theory. Nevertheless, {\em successful models usually open the way to
theoretical advances}. A typical example taken from classical mechanics is
Kepler's model of planetary motion, based on elliptical orbits, 
compared to the standard Copernican model,
based on circular motion and epicycles. Kepler's model was clearly superior in 
predicting the position of planets. This observation opened the way to the
development of 
Newton's theory of gravitation. 

Nowadays, in the field of hadronic physics, we face the following situation:
we have a large (and ever increasing) amount of data, 
but we cannot compare them with first
principle calculations based on QCD. Most of the time, we tend to parametrize
the nonperturbative functions involved in physical observables. 
We are doing something similar to recording the position of planets with high
precision. We cannot offer any interpretation of the data. We cannot establish
connections between different nonperturbative quantities. We have no reason to
trust extrapolations. We are not testing in any way our knowledge of the
structure of the proton. At most, we can test the validity of the framework we
use to analyze the data, i.e., we can test QCD in the perturbative
regime. 

In this situation, resorting to models may be essential. It allows us to
systematize the information we get from different observables and make
predictions based on the assumptions that characterize the model.

\begin{figure}
\begin{center}
\includegraphics[width=0.45 \columnwidth]{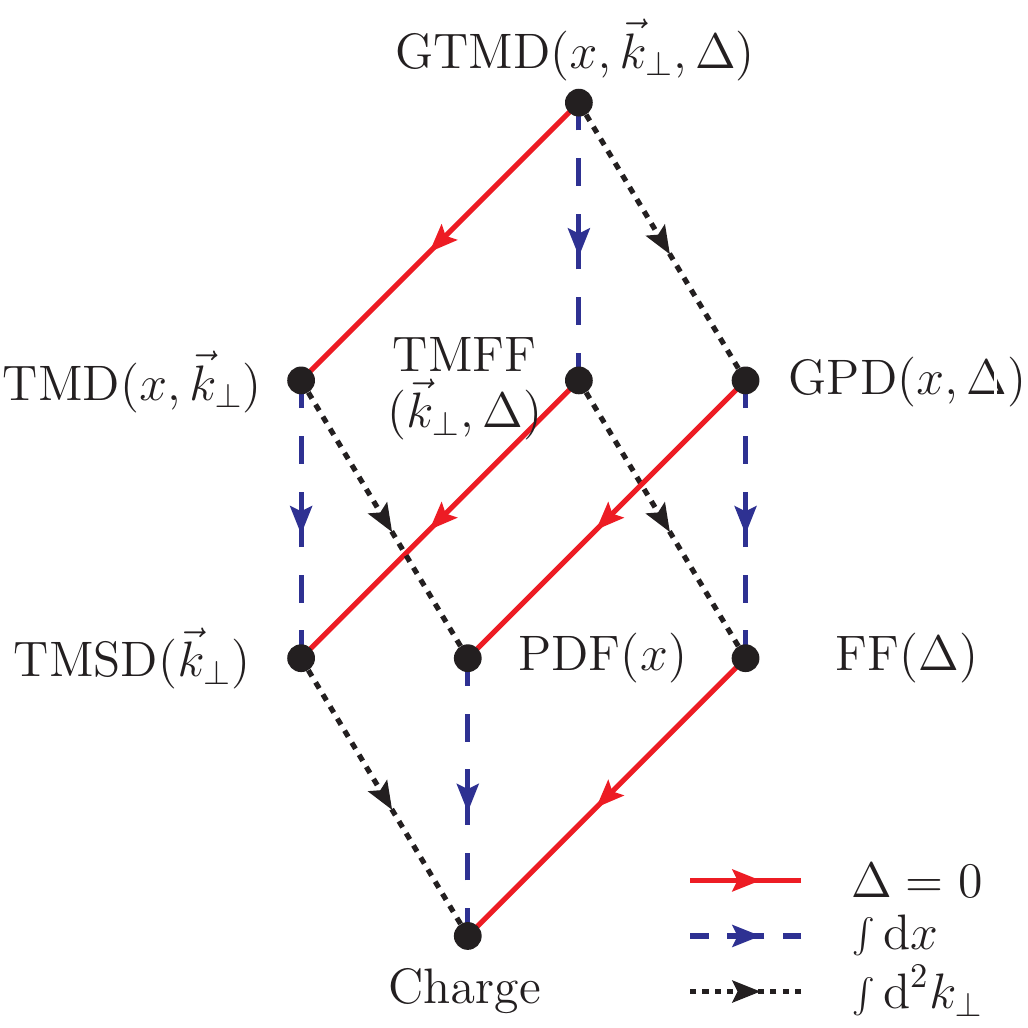}
\end{center}
\caption{Representation of the projections of the GTMDs into 
parton distributions and form factors (picture from ref.~\cite{Lorce:2011dv})}
\label{f:PDcube}
\end{figure}

There is another crucial issue that makes models almost unavoidable when
studying the partonic structure of the nucleon. It is well known that a full
information on the ``phase-space distribution'' of partons can be encoded in
Wigner distribution functions~\cite{Ji:2003ak}. They are Fourier transforms of the so-called
Generalized Transverse-Momentum Distributions (GTMDs)~\cite{Meissner:2009ww}. These objects are
extremely rich and powerful and give us a picture of the partonic structure of
the nucleon in a multidimensional space. Unfortunately, 
they cannot be probed directly. 
In specific limits or after specific
integrations (see Fig.~\ref{f:PDcube}), 
they reduce to Transverse-Momentum Distributions (TMDs) and
Generalized Parton Distributions (GPDs). Further limits/integrations reduce
them to collinear Parton Distribution Functions (PDFs) and Form Factors
(FFs). These four types of partonic distributions can be extracted from
experimental measurements, but often only in limited ranges or limited
combinations of 
the variables involved. In other words, we cannot directly reproduce the full
multidimensional picture
of the inner structure of the nucleon. At most, we can access some particular
projections of the full image. 
We find ourselves as in the Indian
story of the blind men and the elephant (I heard this story first from
H. Avakian):  
six blind men were asked to determine
what an elephant looked like by feeling different parts of the elephant's
body. Each one feels a different part, but only one part. In the end, each one
believes the elephant is something different: a spear (tusk), a fan
(ear), a snake (trunk), and so forth. This story illustrates how misleading
can be to 
draw conclusions based on narrow projections of the whole picture. 

In this brief review, I give for granted TMD definitions, that can be found,
e.g., in refs.~\cite{Bacchetta:2006tn,Boer:2011fh}

\section{Some model results}

Many models have been proposed to describe some aspects of the structure of the
nucleon. However, not many of them have been used to study TMDs and GPDs. In
this talk, I will focus in particular on the following categories of models:
\begin{itemize}
\item Light-cone constituent-quark models (LCCQM)~\cite{Pasquini:2008ax,Pasquini:2010af,Lorce:2011dv}
\item Spectator models (SM)~\cite{Jakob:1997wg,Bacchetta:2003rz,Gamberg:2007wm,Cloet:2007em,Bacchetta:2008af,Bacchetta:2010si}
\item Bag model (BM)~\cite{Yuan:2003wk,Courtoy:2008dn,Avakian:2010br} and
  non-relativistic constituent-quark models~\cite{Courtoy:2008vi,Courtoy:2009pc}.
\item Chiral quark-soliton model ($\chi$QSM)~\cite{Wakamatsu:2000fd,Wakamatsu:2009fn,Lorce:2011dv}
\item Covariant parton
  model (CPM)~\cite{Efremov:2004tz,Zavada:2007ww,Efremov:2009ze,Efremov:2010mt} 
\end{itemize}

In order to judge the reliability of the models, let us first analyze some
well-known functions, i.e., form factors and collinear PDFs. I
will present only some selected results, suitable to convey the general message.

\begin{figure}
\begin{tabular}{ccc}
\includegraphics[height=5.4cm]{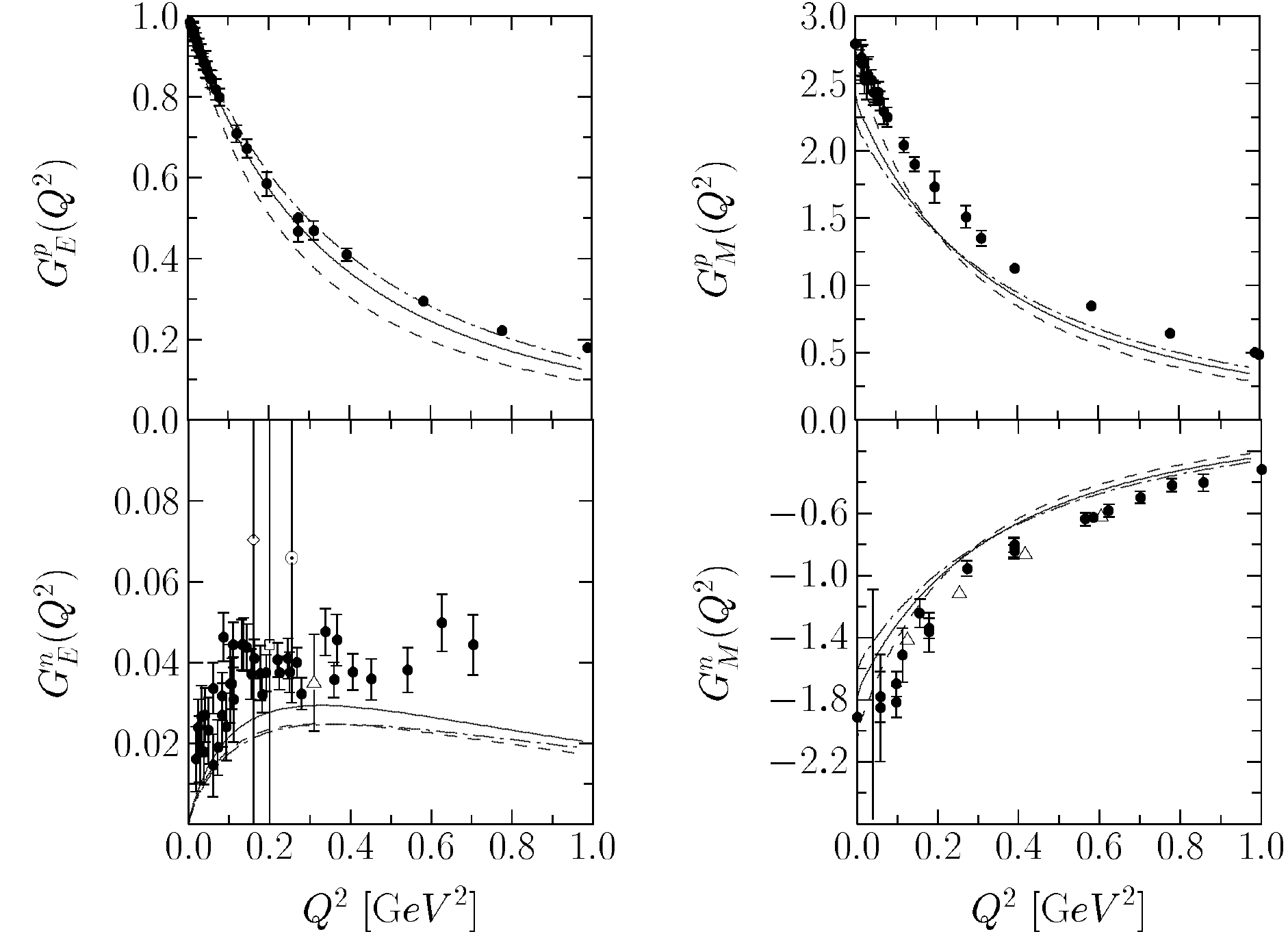}
&
\qquad
&
\includegraphics[height=5.7cm]{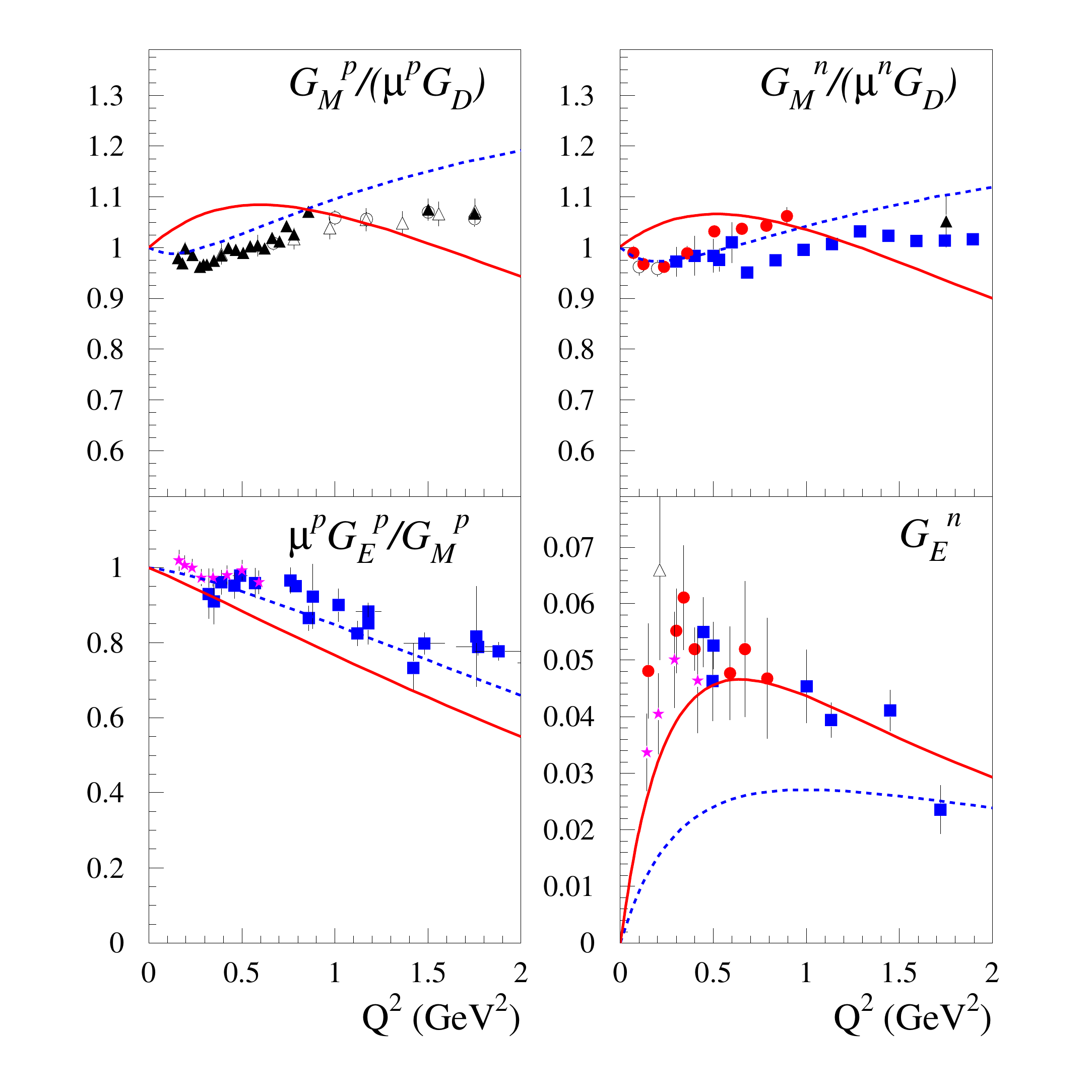}
\\
(a) && (b)
\end{tabular}
\caption{
The four nucleon electromagnetic Sachs FFs compared to 
the world data~(see \cite{Perdrisat:2006hj} and references therein).
(a) Calculation in the $\chi$QSM~\cite{Christov:1995vm} for different values
of the constituent quark mass. 
(b) Solid curves: results in the light-cone $\chi$QSM~\cite{Lorce:2011dv}.
Dashed curves: results from the LCCQM~\cite{Pasquini:2007iz}.}
\label{f:FF}
\end{figure}

In
fig.~\ref{f:FF}, three model
calculations of the proton and neutron electric and magnetic form factors are
compared to available data. 

\begin{figure}
\begin{tabular}{cc}
\includegraphics[height=3cm]{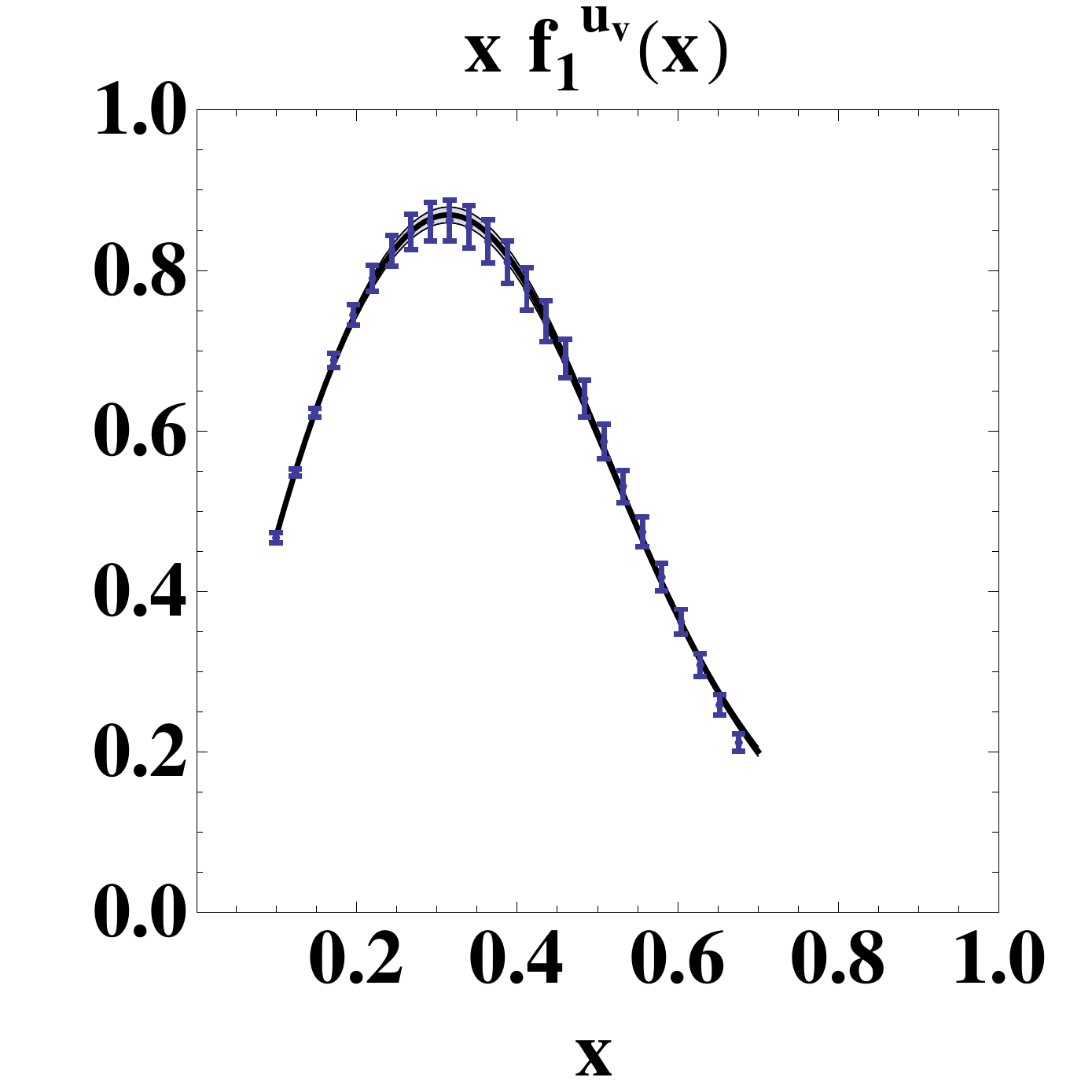}
\hspace{0.1cm}
\includegraphics[height=3cm]{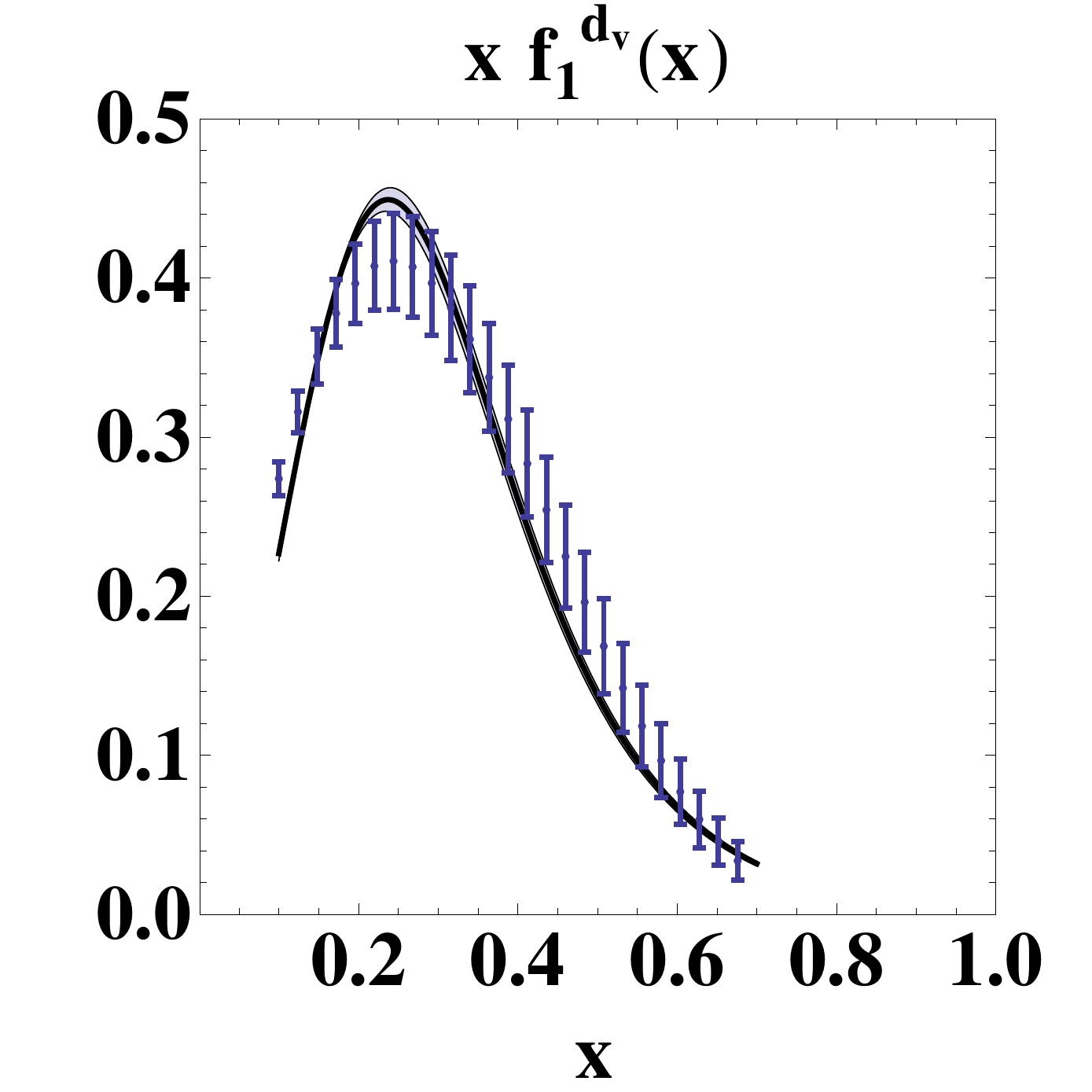}
&
\includegraphics[height=2.6cm]{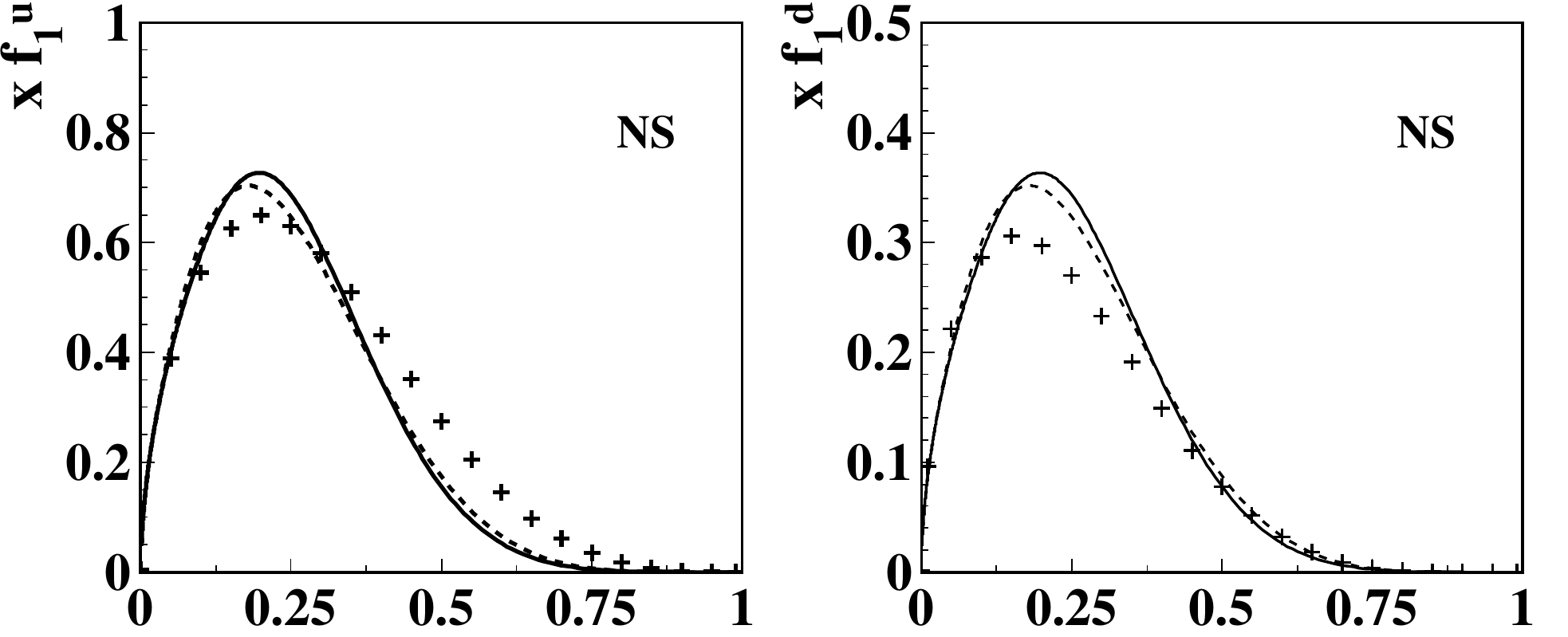}
\\
\includegraphics[height=3.1cm]{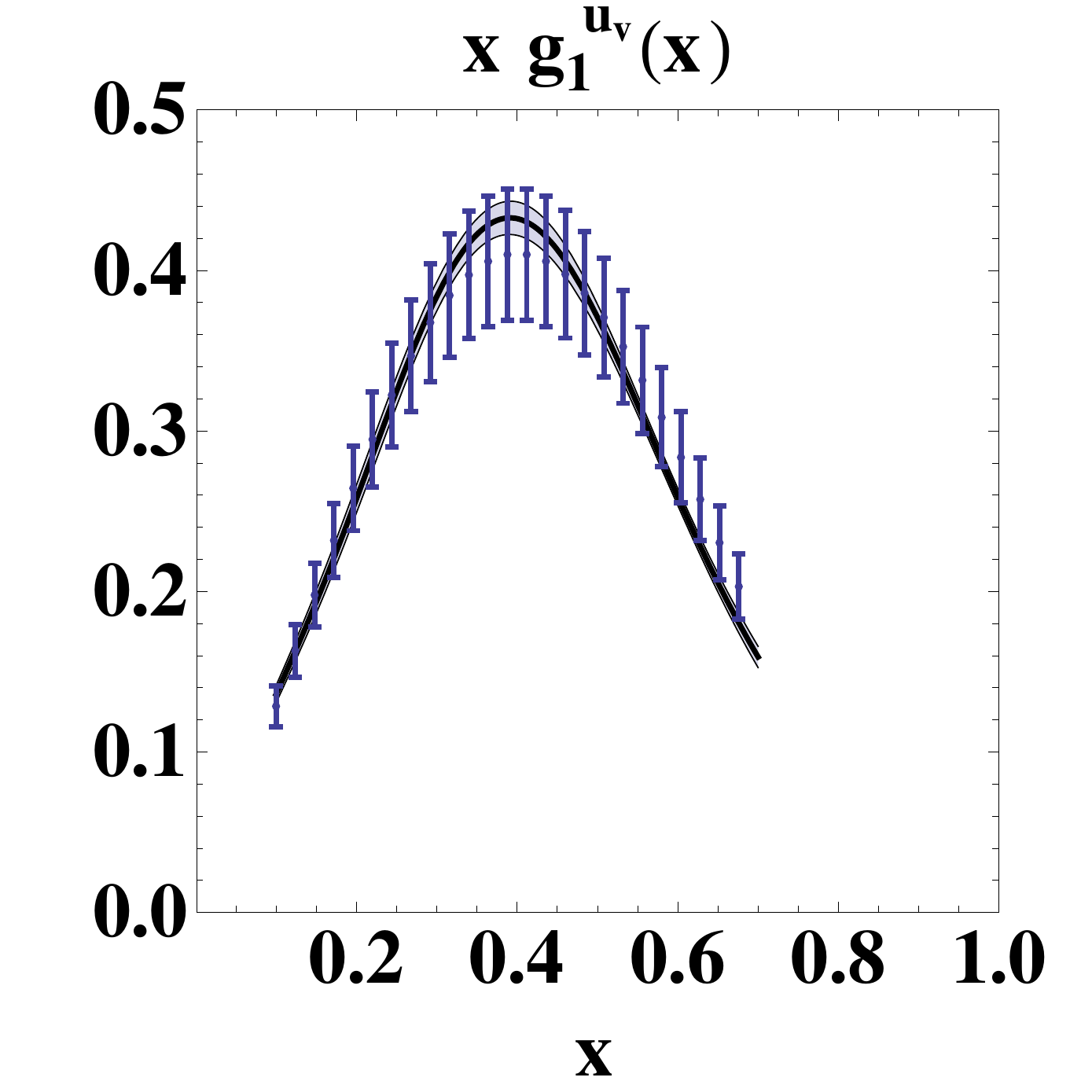}
\includegraphics[height=3.1cm]{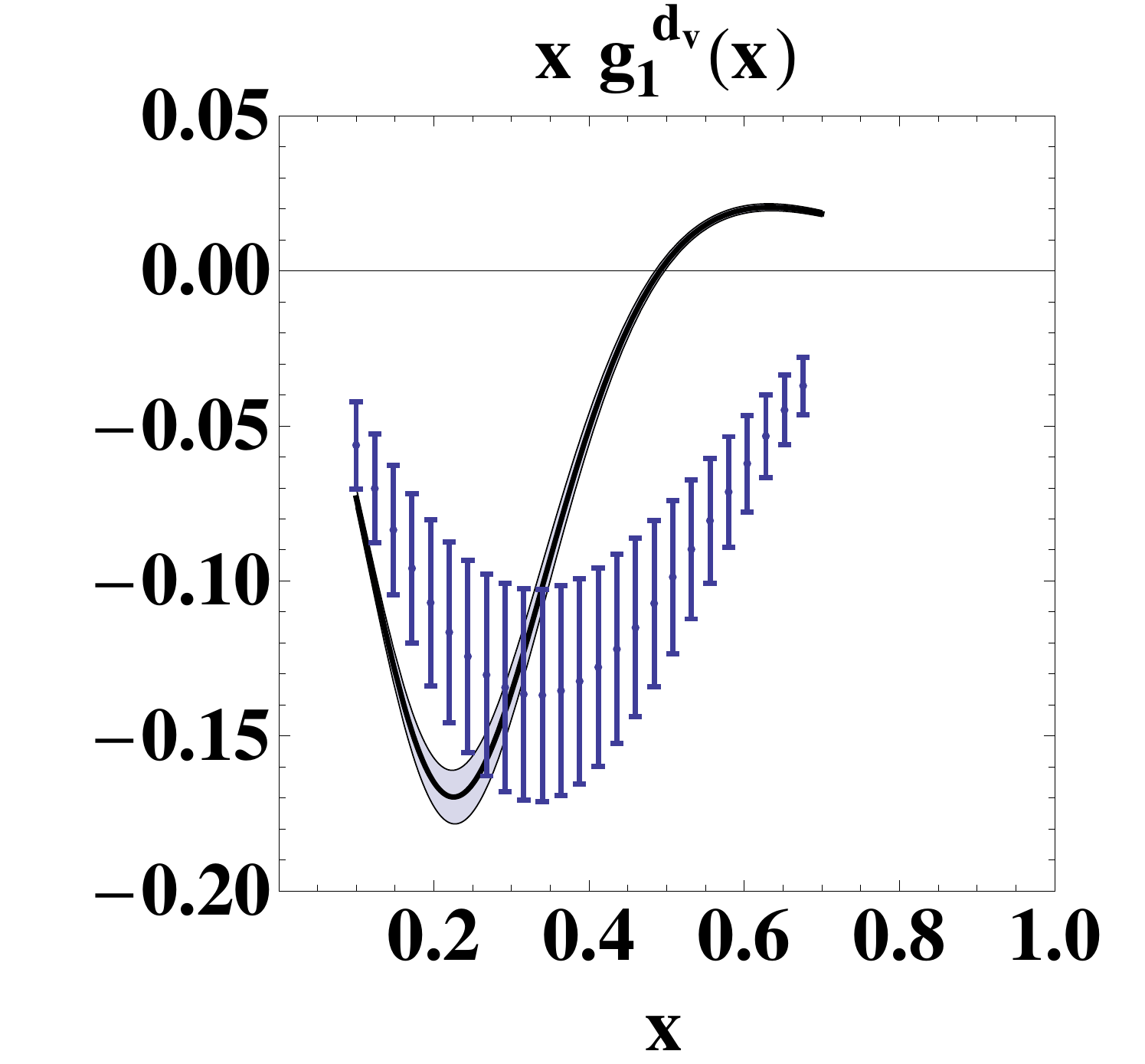}
&
\includegraphics[height=2.6cm]{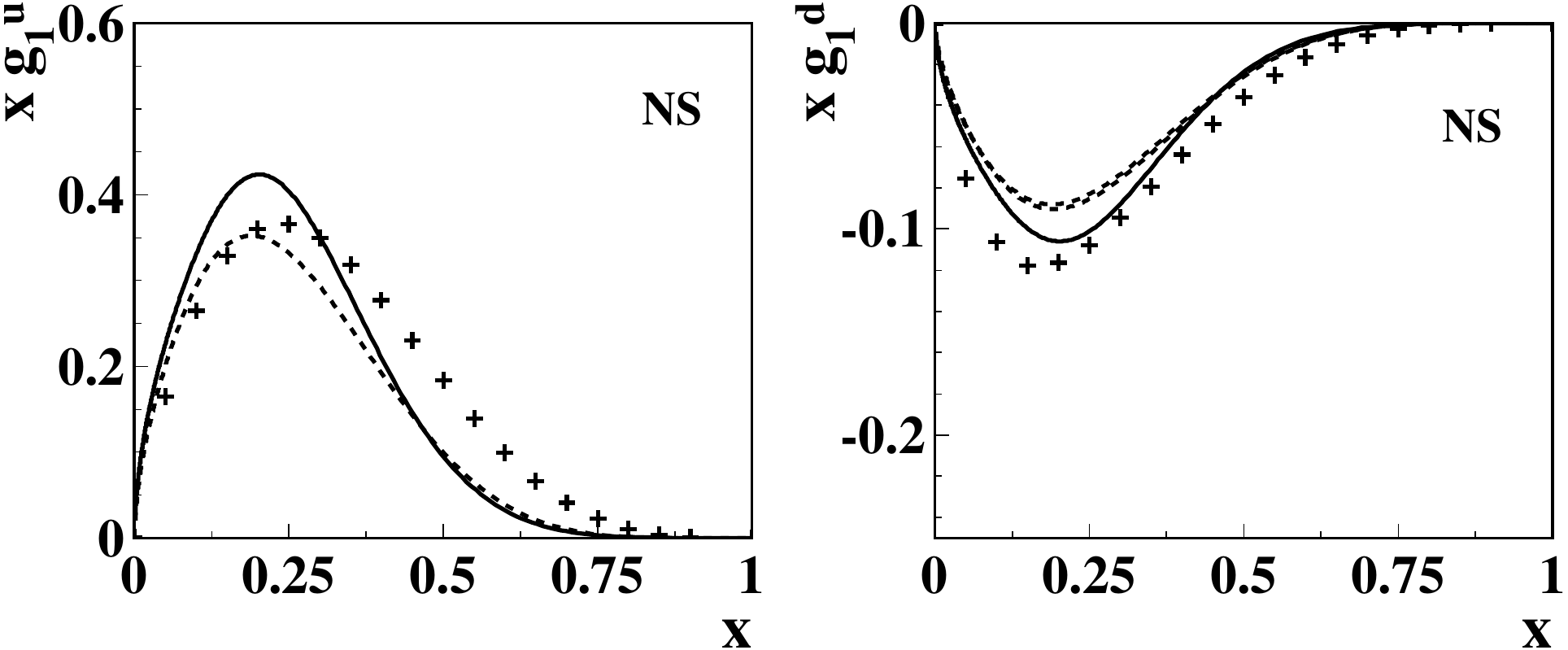}
\\
(a) 
& 
(b)
\end{tabular}
\caption{
The collinear PDFs.
(a) Results in the spectator model~\cite{Bacchetta:2008af}  at 0.3 GeV$^2$
compared to the ZEUS02~\cite{Chekanov:2002pv} and GRSV00~\cite{Gluck:2000dy}
parametrizations. 
(b) Solid curves: results in the $\chi$QSM~\cite{Lorce:2011dv}.
Dashed curves: results from the LCCQM~\cite{Pasquini:2006iv} at 5 GeV$^2$
compared to MSTW09~\cite{Martin:2009iq} and LSS07~\cite{Leader:2006xc} 
parametrizations.
}
\label{f:f1g1}
\end{figure}

In
fig.~\ref{f:f1g1}, three model
calculations of the up and down unpolarized PDF and helicity PDF 
are
compared to available parametrizations. 
I think the general message at this point is the following: models seem to be
able to capture the qualitative features of form-factors and collinear PDFs,
but they are still far from giving a description that is satisfactory from the
quantitative point of view.

\begin{figure}[h]
\begin{center}
\includegraphics[height=5.5cm]{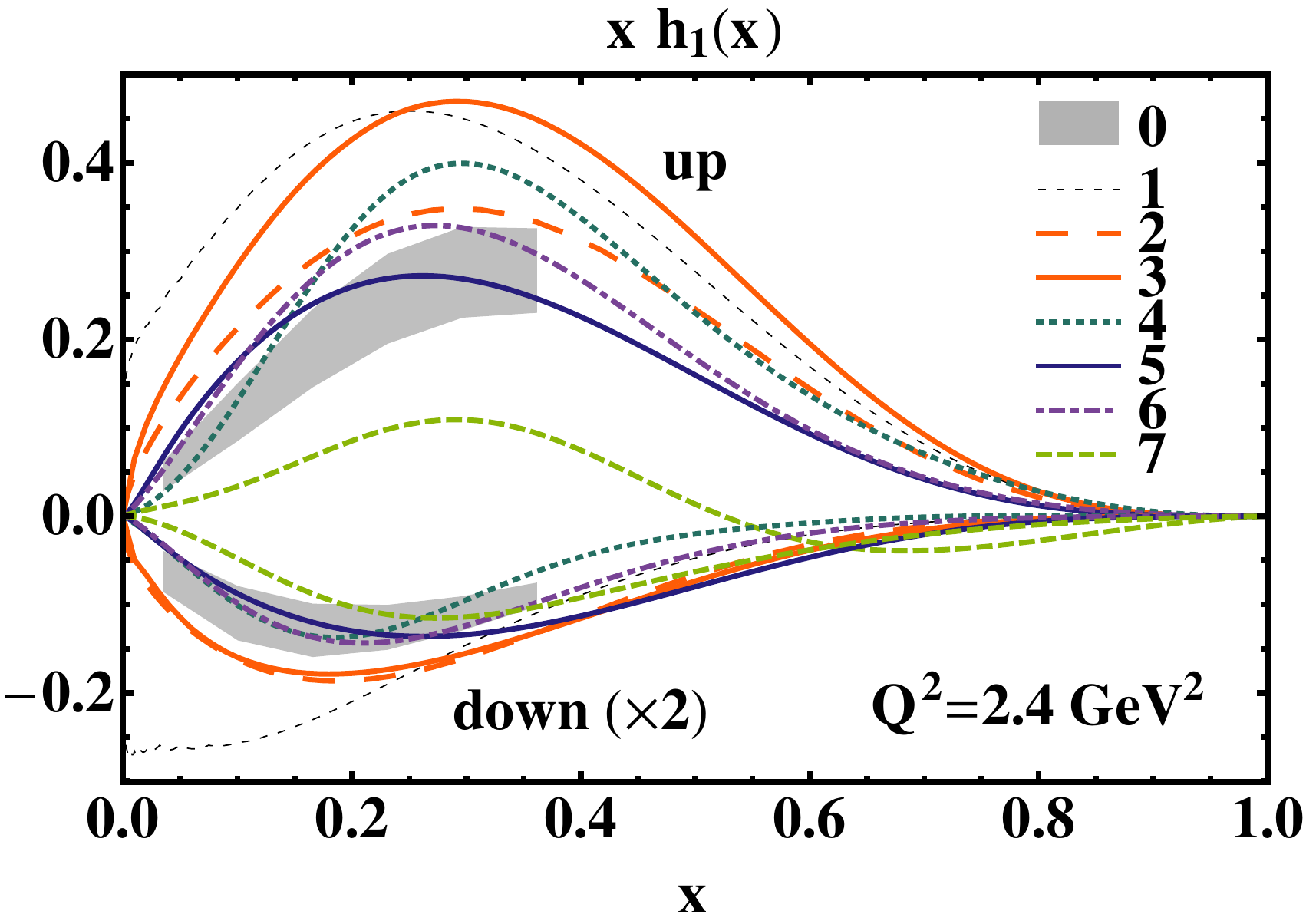} 
\end{center}
\caption{
Model calculations of the transversity distribution function compared to
available parametrizations: (0--shaded band) extraction from ref.~\cite{Anselmino:2008sj}; (1) saturated Soffer
bound~{\cite{Soffer:2002tf}}; (2) $h_1=g_1$~{\cite{Korotkov:1999jx}}; (3-4) chiral
quark-soliton models~{\cite{Schweitzer:2001sr,Wakamatsu:2007nc}}; (5) light-cone
constituent quark 
model~{\cite{Pasquini:2005dk}}; (6-7) quark-diquark 
models~{\cite{Cloet:2007em,Bacchetta:2008af}}.}
\label{f:h1models} 
\end{figure}

Due to the crucial role played at this workshop by
the transversity distribution function, I present in fig.~\ref{f:h1models}
several model calculations of the transversity compared to the presently
available parametrization of ref.~\cite{Anselmino:2008jk}.
The present
extraction of transversity for the up quark is systematically lower than most
model calculations, while for the down quark the extraction tends to be 
bigger than models in absolute value. Two important remarks are in order:
first of all, it must be kept in mind that there are no data at
$x > 0.4$; secondly, it must be stressed that the {\em sign} of the up quark
distribution cannot be determined from experimental measurements but it is
fixed according to the sign of model calculations (interestingly, all of them
give the same sign).

Fig.~\ref{f:h1models} reminds us of another obvious observation: it can often
happen that two predictions based on the same type of model give significantly
different results. In fig.~\ref{f:h1models}, 
this occurs for the two $\chi$QSM predictions
and the two spectator-model predictions.\footnote{Note that the chiral quark-soliton model has a ``model accuracy'' of (10-30)\%
  due to the $1/N_c$ expansion and instanton packing fraction.} 
This is due to different choices in
the implementation of the model or simply to different choices for the
parameters used in the model.

For what concerns TMDs, we should start from the distribution of unpolarized
quarks in an unpolarized nucleon. It has been computed in spectator
models~\cite{Jakob:1997wg,Bacchetta:2008af},  
LCCQ model~\cite{Pasquini:2008ax}, $\chi$QS
model~\cite{Wakamatsu:2009fn,Lorce:2011dv}, bag model~\cite{Avakian:2010br}. 
However, my impression is that at the moment
two important ingredients are missing to allow a fair comparison. From the
phenomenology side, the information on the unpolarized TMDs is still
limited, in particular in the valence region (see a recent analysis in
ref.~\cite{Schweitzer:2010tt}). From the theory side,
model predictions are valid at some low scale, below 1 GeV, 
that is effectively assumed to be the threshold between the nonperturbative
and perturbative regimes. Obviously, comparison with experimental data
requires that the predictions are evolved to a higher scale. This demands the
application of the correct TMD evolution
equations~\cite{Collins:1984kg,Aybat:2011zv}, a task that has not been
carried out yet. The effect of TMD evolution has been shown to be rather
dramatic~\cite{Aybat:2011zv} and tends to broaden the transverse-momentum
distribution when the scale increases.

In any case, let me present some numerical results of the
models. At a scale of about 0.1 GeV$^2$, 
the LCCQ model of ref.~\cite{Pasquini:2008ax,Boffi:2009sh} gives a 
$\langle k_\perp^2 \rangle=0.080$ GeV$^2$, and the bag model of
ref.~\cite{Avakian:2010br} gives $\langle k_\perp^2 \rangle=0.077$ GeV$^2$ (at $x=0.3$). In
these two models, the
width is predicted to be the same for up and down quarks (valence only). 
The spectator model of ref.~\cite{Bacchetta:2008af}
gives larger widths and different for the two flavors: 
$\langle k_{\perp u_v}^{2} \rangle=0.87$ GeV$^2$ and
$\langle k_{\perp d_v}^{2} \rangle=0.16$ GeV$^2$
at a
scale of 0.30 GeV$^2$. Finally, the $\chi$QSM of ref.~\cite{Wakamatsu:2009fn}
predicts $\langle k_{\perp (u+d)}^{2} \rangle=0.18$ GeV$^2$ and 
$\langle k_{\perp (\bar{u}+\bar{d})}^{2} \rangle=0.62$ GeV$^2$ at a
scale of 0.30-0.40 GeV$^2$.

There is a conceptual problem when comparing model TMDs and PDFs to
extractions: formally---at least in the present formulation of TMD
factorization---collinear PDFs are {\em not} the integrals of the
corresponding TMDs~\cite{Collins:1984kg,Aybat:2011zv}. The two parton
distributions are connected in a more complex way. This may be intuitively
understood as follows: the TMD formalism is applicable only when the
transverse momentum is much smaller than the dominant hard scale; therefore,
after integration it cannot fully reproduce collinear PDFs, 
which entail an integration over
transverse momentum up to a value of the order of the hard scale.

\begin{figure}
\begin{center}
\includegraphics[height=5cm]{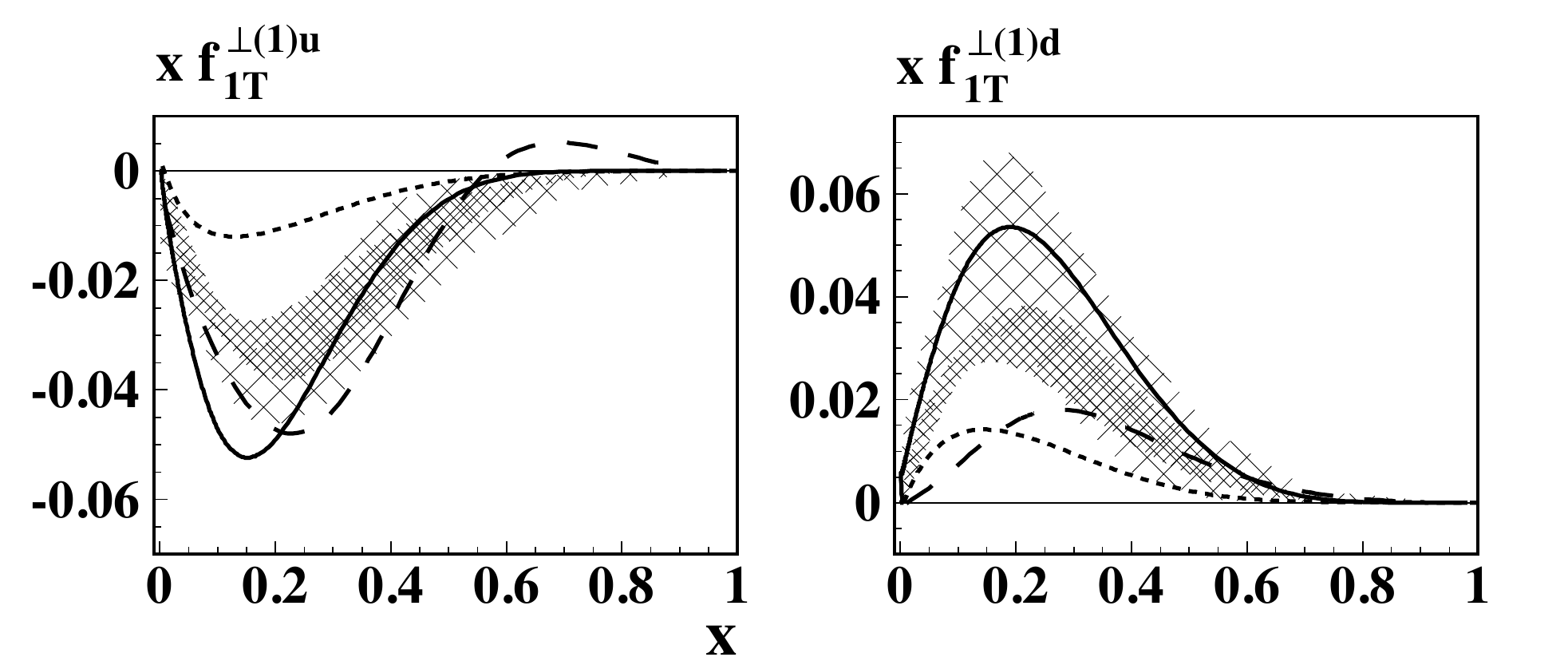} 
\end{center}
\caption{Results for the (1)-moments
of the quark Sivers function. Solid curves: light-cone constituent
quark model of ref.~\cite{Pasquini:2010af}. Dashed curves:
spectator model of ref.~\cite{Bacchetta:2008af}. Dotted
curves: bag model of ref.~\cite{Courtoy:2008dn,Courtoy:2009pc}.
Shaded bands correspond to the
parametrizations 
of ref.~\cite{Anselmino:2008sga} and
\cite{Efremov:2004tp,Collins:2005ie}.}
\label{f:sivers_models} 
\end{figure}

For the moment, let us neglect these problems and turn our attention
to another important TMD: the Sivers function~\cite{Sivers:1990cc}. At
present, this is the only TMD for which we have a few consistent
extractions from experimental
data~\cite{Efremov:2004tp,Vogelsang:2005cs,Collins:2005ie,
Anselmino:2008sga,Arnold:2008ap,Anselmino:2011gs,Bacchetta:2011gx}. There is
also a long list of model calculations of the Sivers function in the last
decade: in spectator
models~\cite{Brodsky:2002cx,
Gamberg:2003ey,Goldstein:2002vv,Bacchetta:2003rz,Lu:2004au,Goeke:2006ef,
Lu:2006kt,Ellis:2008in,Bacchetta:2008af},  
in the bag
model~\cite{Courtoy:2008dn,Yuan:2003wk,Courtoy:2009pc},
in a non-relativistic constituent quark model~\cite{Courtoy:2008vi} and
a light-cone constituent quark model~\cite{Pasquini:2010af}.  In
fig.~\ref{f:sivers_models}~\cite{Boer:2011fh}, the results of three different
calculations of the 
Sivers function~\cite{Pasquini:2010af,Bacchetta:2008af,Courtoy:2009pc} are
compared to two of the parametrizations.  
By comparing models with extractions, 
we can appreciate the fact that models obtain the
right sign for the up and down Sivers functions and the right order of
magnitude. However, there are large discrepancies in the size. Once again, if
we aim at using model for quantitative studies, we need to improve them.

\section{Relations among TMDs in models}

Model calculations typically have a specific structure and a set of
parameters for numerical predictions. If a model disagrees with experimental
observations, it may be due to the structure of the model being wrong, or more
trivially to the wrong choice of parameters.
A powerful way to investigate the structure of the model is to search for
relations among predicted quantities, independent of the specific choice of
parameters. Several works addressed this question in the last
years~\cite{Avakian:2007mv,Metz:2008ib,Lorce:2011zt}. 
Relations based on models have been established
among TMDs, between TMDs and PDFs~\cite{Efremov:2010mt}, and between TMDs and
GPDs~\cite{Burkardt:2003je,Meissner:2007rx}. 

A commonly used set of relations is represented by the so-called
Wandzura--Wilczek 
relations~\cite{Wandzura:1977qf}. These relations involve the so-called
twist-3 partonic distributions, which typically occur in experimental
observables suppressed by a power $M/Q$. At the partonic level, ``pure
twist-3'' usually refers to quark-gluon-quark correlations. The
Wandzura--Wilczek approximation consists in removing all pure twist-3
terms. This is strictly speaking possible only for a non-interacting theory, of
the kind suggested for instance in the covariant parton
model~\cite{Zavada:2002uz} and in ref.~\cite{D'Alesio:2009kv}. In any theory
with interactions the Wandzura--Wilczek relations should
break. The group of 
T-odd TMDs
requires final-state interactions and therefore vanishes in
Wandzura--Wilczek approximation.

There is no strong 
evidence yet of the violation of the Wandzura--Wilczek relations. Indications 
can be found
in the structure function $g_2$ (see, e.g., ref.~\cite{Accardi:2009au}) as
well as in the fact that the T-odd Sivers function is
nonzero~\cite{Airapetian:2004tw,Airapetian:2009ti,Alekseev:2010rw}. It may
nevertheless still be possible to invoke these relations to make
semi-quantitative estimates, in particular for T-even
TMDs~\cite{Avakian:2007mv}. 

A second class of relations is represented by the so-called Lorentz-invariance
relations. This is actually a misnomer because they are not descending
directly from Lorentz invariance. Rather, they originate from neglecting a
certain subset of pure twist-3 contributions, namely those related to gauge
fields~\cite{Goeke:2003az,Accardi:2009au}. As for the
Wandzura--Wilczek relations, T-odd functions should be zero within these
assumptions. 
Models without gluons respect these relations,
while they are obviously spoiled by perturbative QCD, as shown explicitly by 
the quark-target model~\cite{Kundu:2001pk,Accardi:2009au}. 

A third class of relations has been observed in a few model calculations (see,
e.g., 
\cite{Avakian:2009jt} and references therein)
and discussed in depth in
ref.~\cite{Lorce:2011zt}. They read 
\begin{gather}
g_{1}-h_1 = \tfrac{k_\perp^2}{2M^2}\,h_{1T}^{\perp},\label{rel1}\\
g_{1T}=-h_{1L}^{\perp},\label{rel2}\\
f_{1T}^\perp = h_1^{\perp}=0, \\
\left(g_{1T}\right)^2+2h_1\,h_{1T}^{\perp}=0.\label{rel3}
\end{gather} 

These relations are ascribed to the spherical
symmetry of the partonic wave-functions in the canonical-spin basis (together
with the possibility of rotating from canonical spin to light-cone
spin). They appear to be violated in models with spin-one particles
(gluons or axial-vector diquarks). They are violated in perturbative
QCD. Nevertheless, they could still be valid approximations at the threshold
between the nonperturbative and perturbative regimes. As such, they could be
useful conditions to guide TMD parametrizations.

\section{Angular momentum and TMDs in models}

In general, 
it is not possible to get direct access to partonic angular momentum using
TMDs. This is however possible in the context of models, since they imply
characteristic relations between formally independent functions. 

A first example has been discussed in a version of the spectator
model~\cite{She:2009jq}, in the bag model~\cite{Avakian:2010br} and in the
covariant parton model~\cite{Efremov:2010cy} and can be written as
\begin{equation}
{\cal L}_z^q = - \int dx h_{1T}^{\perp (1) q} (x) = 
\int dx \Bigl(h_1^q (x) -g_1^q (x)\Bigr).
\end{equation} 
In the discussion of this relation it has been observed that the identity is
valid only at the level of matrix elements, not of operators. 
If the contributions of all partons
are summed, the definition
of the orbital angular momentum on the l.h.s.\ corresponds to that of Jaffe--Manohar.  
Since
the relation has been discussed in a theory without gauge fields, there should
be no distinction between the definition of orbital angular momentum according to
Jaffe--Manohar and Ji (see, e.g., the discussion in
refs.~\cite{Burkardt:2008ua,Wakamatsu:2010cb}).

A different connection has been explored in
ref.~\cite{Bacchetta:2011gx}. Based on results of spectator 
models~\cite{Burkardt:2003je,Lu:2006kt,Meissner:2007rx,
  Bacchetta:2008af,Bacchetta:2010si} 
and theoretical considerations~\cite{Burkardt:2002ks},  
the following relation
has been used
 \begin{equation}
 f_{1T}^{\perp (0) q}(x) = -L(x)\, E^q (x,0,0),
\label{e:EtoSivers0}
\end{equation}  
connecting the Sivers TMD with the GPD $E$ in the forward limit, which is
essential to compute angular momentum using Ji's
relation~\cite{Ji:1997ek}. 
The above relation can be written in a more general form as a convolution between the ``lensing function'' $L$ and the GPD
$E$. It is actually one of many relations that can be established between TMD
and GPDs, which are however also model-dependent~\cite{Meissner:2007rx}.

In ref.~\cite{Bacchetta:2011gx}, relation \eqref{e:EtoSivers0} was assumed to
be valid and it was shown that it is possible to fit at the same
time the nucleon anomalous magnetic moments and data for semi-inclusive
single-spin asymmetries produced by the Sivers effect. This analysis leads to
the following estimate for quark angular momenta (at $Q^2 = 4$ GeV$^2$)
\begin{align} 
J^u &= 0.229 \pm 0.002^{+0.008}_{-0.012} ,
&
J^{\bar{u}} &= 0.015 \pm 0.003^{+0.001}_{-0.000} ,  
\\
J^d &= -0.007 \pm 0.003^{+0.020}_{-0.005} ,
&
J^{\bar{d}} &= 0.022 \pm 0.005^{+0.001}_{-0.000} .  
\end{align} 
This represents the first model-inspired extraction of partonic angular
momentum from TMDs. Strikingly, the results are in agreement with other
totally independent estimates~\cite{Diehl:2004cx,Guidal:2004nd,Ahmad:2006gn,Goloskokov:2008ib,Wakamatsu:2007ar,Bratt:2010jn}.

\section{Conclusions}

A wealth of results concerning transversity and TMDs 
has been obtained from model calculations in the last decade. Apart from
making useful predictions on yet unmeasured quantities, they help us to
better understand experimental results. Since they provide relations between
different partonic distribution functions, they provide hints on how to
reconstruct the full multidimensional structure of the nucleon starting from
a limited number of ``projections.''

\acknowledgments
I thank C\'edric Lorc\'e, Barbara Pasquini, Marco Radici, and Peter Schweitzer for useful
discussions on the 
subject. 
This work is partially supported by the Italian MIUR through the PRIN
2008EKLACK. 

\bibliographystyle{myvarenna}
\bibliography{mybiblio}

\end{document}